# Future-Proofing IoT: Unleashing the Power of AWS Greengrass in Propelling Smart Devices to New Heights

K Sahasra
Student
*Networking and Communication*
*SRM Institute of Science and Technology, Kattankulathur*
Chennai, India
kk4785@srmist.edu.in

Ankit Vatsa
Student
*Networking and Communication*
*SRM Institute of Science and Technology, Kattankulathur*
Chennai, India
as6553@srmist.edu.in

Dr.P.Savaridassan
Assistant Professor
*Networking and Communication*
*SRM Institute of Science and Technology, Kattankulathur*
Chennai, India
savaridp@srmist.edu.in

*Abstract*— The advent of edge computing is set to revolutionize cloud computing in various sectors, including Agriculture, Health, and more. AWS Greengrass Core Device plays a pivotal role in this transformative process by bridging connections between IoT devices by improving data sharing between them, unlocking new possibilities in Smart Home, Agriculture, Health, Vehicular Cloud, Smart City, Industry Automation, and beyond. However, these advancements also introduce novel challenges for testing and quality assurance in cloud computing. This paper explores the impact of AWS Greengrass in different fields, addressing challenges, opportunities, and potential benefits of edge and cloud computing in terms of processing speed, latency, and bandwidth usage.

## 1. INTRODUCTION

One of the most significant recent technologies is edge computing. Organizations can take use of data-consuming technologies like artificial intelligence (AI), biometrics and surveillance the Internet of Things, and endpoint automation owing to the edge.[22]

When Akamai introduced its content delivery network (CDN) in the 1990s, edge computing first appeared. during that time, the plan was to place nodes closer to the user's location to deliver cached information like photographs and videos.[23]

The use of edge computing was hastened by the expansion of Internet of Things (also known as IoT) devices. As the number of linked devices increased quickly, it became impossible and inefficient to handle all the data produced by these machines in a single location. Edge computing provided a remedy by moving processing closer to the actual devices. As a result, less data must be transferred to a centralized server or cloud, reducing latency and increasing overall system efficiency for IoT devices.[25]

A piece of programming called Greengrass gives local devices access to cloud services. Devices may now collect and evaluate data more locally, react to local events on their own, and securely communicate via local networks. Local devices can securely interact with AWS IoT Core and publish IoT data to the AWS Cloud. [24]

More "things" will be networked to the internet as the cost of sensors and computers continues to drop. Additionally, edge computing will find additional uses across industries as additional devices with connectivity become available, especially when cloud computing occasionally proves to be ineffective. Its effects are already starting to be felt across various industries. Here are a few industries that could profit from edge computing, from driverless vehicles to agriculture.[26]

### 1.2 ROLE IN INDUSTRIES

As IoT and cloud technologies become more familiar, industries of all sizes are adopting IoT for streamlined machine functioning. Real-time data and statistics are transferred to their dashboards, facilitating easy maintenance and giving engineers greater control. With data collection from various sensors made simple, AI-based monitoring at the edge ensures efficient and prompt machine operations.

### 1.3 ROLE IN TRANSPORTATION

IoT, 5G technology and automation are set to greatly impact the transportation field by enabling new use cases such as driverless cars and vehicular clouds. These solutions can reduce traffic congestion and improve efficiency significantly. For example, driverless cars rely on advanced sensors and AI to navigate and avoid obstacles, reducing the chance of human error that causes much traffic congestion. V-clouds are a new way of utilizing resources where vehicles can communicate with each other and form self-organizing networks to collect data and conduct computation.

## 1.4 SMART CITY

As technology continues to advance, more and more devices, from small to large, are incorporating artificial intelligence to work intelligently. As the idea of Digital India came into picture, the concept of smart cities has emerged, which involve the use of electronic methods and sensors to gather data, and using the information gained from that data to efficiently manage resources, assets, and services in urban areas.

Video surveillance and traffic management are some of the crucial needs for a smart city. Surveillance includes scanning license plates of vehicles and analysis of any violence-related activity with the help of artificial intelligence and live traffic data for intelligent traffic management.

## 1.5 E-HEALTH

For the efficient and secure use of technology in the field of healthcare, including services, surveillance, literature, education, knowledge, and research E-health concept arises to help citizen.

## 1.6 SMART HOME

In this tech-driven world, remote home monitoring with smart video surveillance systems and cameras grants convenience and control from anywhere. IoT optimizes appliance usage, allowing intelligent switching on and off based on usage patterns, while also providing real-time notifications for important events. Moreover, smart kitchen devices, like coffee machines, offer remote operation for added ease and comfort.

## 2. LITERATURE STUDY

The main objective of this paper is to examine the different approaches presented by the authors so far and analyze their advantages and shortcomings. Around $208 billion, or 13.1% more than 2022, is projected to be spent on edge computing globally in 2023. According to the Worldwide Edge Spending Guide published by International Data Corporation (IDC), enterprises and service providers are expected to continue spending at this rate of increase through 2026, when they expect to have spent close to $317 billion on IT services, software, and hardware for edge solutions[2]. By effortlessly extending AWS to edge devices, Greengrass enables them to respond regionally on the information they produce while continuing to use the cloud for administration, analytics, and long-term retention. Even when not online, linked devices can use AWS IoT Greengrass to run Lambda functions in AWS , make predictions using models trained with machine learning, keep device information synchronized, and also securely interact with one another[3]. Using AWS Greengrass as our edge computer node, edge computing in this case refers to relocating your computing capability and processors as near to your devices and systems as possible[1]. Today's world Greengrass can be used to incorporate edge technology in numerous industries, including manufacturing, energy, healthcare, transportation, retail, agriculture, and gas, logistics and supply chain, and environmental monitoring.

A fresh computing paradigm is necessary as real-time data processing requirements grow as a result of recent application advances. Edge computing overcomes this problem by bringing computing resources needed by Internet of Vehicles and other intelligent transportation systems from the cloud into proximity of the end devices to enhance performance[4]. Vantage Power, a hybrid and electric technology manufacturer, collaborated with Luxoft to create VPVision, a telemetry platform for medium- and heavy-duty commercial vehicles. Luxoft's AWS IoT Analytics and Greengrass ML provide insights into over 6,000 data points from each vehicle. This enables Vantage Power to build and train machine learning models on AWS, enabling faster idea development and testing[5].

Healthcare organizations can handle issues with processing information, security, and real-time monitoring with the help of AWS Greengrass. For example AWS Greengrass and the American multinational company Qualcomm work together to enable both regional and cloud-based data storage for vital Internet of Things applications including industrial machinery and medical equipment[6].

Appliance makers leverage AWS Services to construct and maintain their smart home solutions. The use of Greengrass, a crucial service among them, enables edge devices to have local computation, communication, organising data, synchronise, and ML inference abilities[15].

Important use cases for Industry 4.0 or automated production require access to the sensor feed or diagnostics of industrial gear. Predictive maintenance or automated quality control, for instance, cannot be carried out without such high temporal resolution data. It is challenging to obtain machine data in the diverse setting of modern industrial production lines with a wide variety of machine types because the machine interfaces may employ many protocols. You can create a general design template for implementing protocol transitions with AWS IoT Greengrass. Providing Lambda services to the AWS IoT Greengrass Core is a prerequisite for transforming protocols using AWS IoT Greengrass. [16].

Rapid technological developments in wireless sensor networks, information technology, and interactions between humans and machines all contribute to the rapid evolution of smart cities. Urban computing offers the processing power necessary to incorporate such technology and raise urban residents' standard of living[17].City data is gathered from an extensive and expanding range of sources. A trained machine learning model that has been implemented with AWS Greengrass can be used to analyze video and audio sources at the edge[18].

With all the inferences made thus far, it is safe to say that AWS Greengrass as a whole has enormous potential that, when unlocked with the proper resources, might lead to the creation of much more effective, robust, fast, and well-equipped technologies, revolutionizing the current IT industry.

## 3. LIMITATIONS

Every new cutting edge technology has its own set of problems, despite the fact that it offers numerous new features that make machines smarter and life easier. For AWS Greengrass, some of these are:

- Despite being an extensive and advanced tool for developing IoT apps, Amazon IoT Greengrass falls short when it comes to environment disc space, offering just approximately 512MB.[21]
- The degree of difficulty: Establishing and managing AWS Greengrass can be challenging, particularly for users who are unfamiliar with the system[20].
- Restricted Equipment Assistance: AWS Greengrass provides support for a select group of devices, which may not be appropriate for all applications[20].
- Another significant disadvantage of AWS Greengrass is the requests severed (Payload 6 MB).
- Insufficient Personalization: Due to AWS Greengrass's limited modification possibilities, it can be challenging to modify your application to specifically cater to the demands of your consumers.[20]
- Restricted Offline Features: AWS Greengrass has restricted offline abilities which means that if the device in question has no connection to the internet, your application might not function correctly.[20]
- Reliance on AWS Services: Because AWS Greengrass is closely connected with other services offered by AWS, using Greengrass effectively requires that you also utilise other AWS services. If you're trying to find a more self-sufficient answer, this may pose a problem.[20]
- Initial installation and deployment of Greengrass devices have occasionally been difficult for some customers, especially when administering a large number of devices. The absence of simpler automated deployment procedures has frequently been demanded.

Table I: Comparison among present edge software [7]

| Edge Software | Real-time response | ML | IoT Core Support | In-built SDK | Open Source | Cloud Agnostic | Enterprise Solution |
|---|---|---|---|---|---|---|---|
| AWS Greengrass | Yes | Yes | Yes | No | No | No | Yes |
| Cisco Fog Director | Yes | No | Yes | No | No | No | Yes |
| Cisco IOx | Yes | No | Yes | Yes | No | Yes | No |
| Foghorn.io | Yes | Yes | Yes | Yes | No | Yes | No |
| Crosser.io | Yes | Yes | No | No | No | No | Yes |
| Swim.ai | Yes | Yes | No | No | No | No | No |
| Macchina.io | Yes | No | Yes | Yes | No | No | Yes |
| EdgeX FoundryTM | Yes | No | Yes | No | Yes | Yes | Yes |
| Apache Edgent | Yes | No | Yes | Yes | Yes | Yes | No |

## 4. IMPACT ON DIFFERNET FIELDS & SECTORS

AWS Greengrass plays a vital role by offering versatile services and functionalities, supporting MQTT and Modbus protocols, making it compatible with a wide range of machines, sensors and actuators. It prioritizes device-friendliness and effectively fulfils the objectives of edge computing to tackle challenges related to minimizing internetwork bandwidth and reducing service latency. This user and device-friendly service makes AWS Greengrass highly adaptable for use in various fields and sectors. Its suitability for various equipment & sensors enables seamless integration across industries, empowering businesses to leverage the benefits of edge computing effectively.[7]

According to data from enlyft, AWS Greengrass is most often used by companies with more than 10,000 employees and revenues exceeding $1000 million When analyzing AWS Greengrass clients by sector, the biggest segments are Semiconductors (5%), IT Equipment (5%), software for computing (5%), Electrical/Electronic Production (10%), and IT Products and Services (15%). [27]

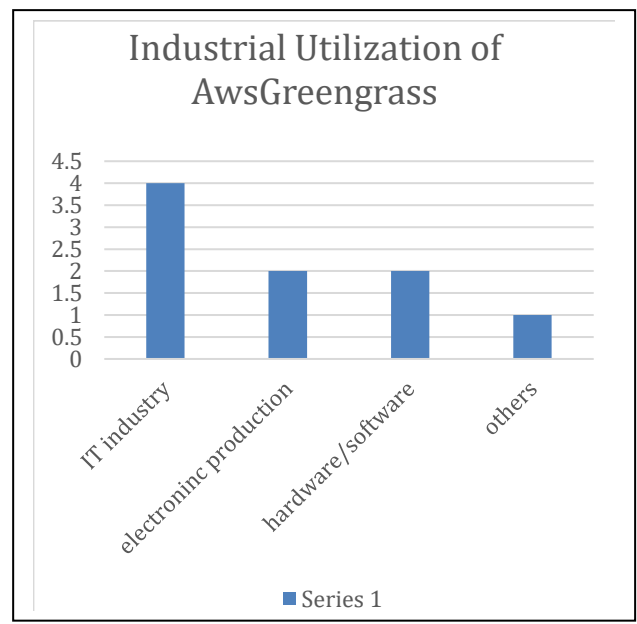

Figure 1. AWS Greengrass in Different Industries (Based on the data published by enlyft[27]).

It finds a seamless integration in diverse domains and fields, ranging from small and large-scale industries to the health sector, agriculture, smart urban citizen applications, infrastructure management, vehicular technology, machine learning inference, and smart homes by enabling enhanced security, efficiency and real-time data processing.[7][8]

Industries, both small and large scale, are increasingly transitioning from off-the-shelf solutions—comprising set of adaptors and scripting languages facilitating connectivity and data forwarding to prebuilt cloud environments dashboard towards framework-based solutions. These frameworks empower businesses to custom-build solutions tailored to their specific needs, with the added advantage of exploring possibilities at the edge, even for complex requirements.[1]

AWS Greengrass Core Device by serving as a bridge between various connected IoT devices plays a crucial role in this transformation. It facilitates the deployment of multiple components (local processes) to communicate with factory machineries, controllers, and sensors while providing intelligent commands to optimize operations effectively.[1][14]

One of the biggest industries that generates the majority of the food needed worldwide is industrial food manufacturing. AWS Greengrass, an infrastructure-independent IoT hardware, will play a significant role in advanced quality check methods, to ensure appropriate quality control and efficient food management, resulting in reduced waste and cost-effectiveness. The Greengrass core device is connected to a number of sensors in order to collect process information and send it to the cloud using wireless communication technologies. This includes an IoT-based tracking and positioning system for materials, items, or equipment using RFID technology, as well as monitoring to minimize waste, improve distribution, and enhance transportation within the fresh food supply chain. As a further approach to digital food manufacturing, an AI-enabled analysis of malt buildings with adjustable moisture, temperature, acidity, and $CO_2$ levels is described. Greengrass core device works collaboratively with these sensors and actuators to command them in appropriate directions at the edge and send relevant pieces of information to the cloud. Also, in the retail sector, it aids in the seamless management of inventory and enables smart automation within stores by swiftly analyzing customer behavior and preferences. [10]

Greengrass, with its fast decision-making, quick response, and high throughput capabilities, proves to be an excellent fit for e-health applications. The device efficiently collects data from connected hardware components such as body temperature, heart rate sensors and wearable healthcare devices. It performs various filtering, aggregation, and dissemination tasks for the collected raw data. This data can either be sent to the IoT cloud via the IoT gateway or processed locally, conserving bandwidth and ensuring low latency. For instance, when a patient's body temperature is normal, the edge device processes the sensor readings locally without uploading them to the cloud. Only when the readings exceed the specified range, the temperature data is uploaded. [7][11]

Agriculture can be intelligently revolutionized with the implementation of Greengrass, enabling automated detection and recognition of key growth stages in vegetables. This allows for precise adjustments in nutrition, care, and environmental conditions to maximize crop yields. Using the camera ecosystem, algorithms for ML are used to identify different stages of plant development based on factors such as length, leaf count, number of blooms, and harvests. The hydroponic agriculture treatment patterns, duration, humidity, and temperature are among the plant care elements that these algorithms ascribe. Greenhouse farming may become more environmentally and economically feasible by streamlining operations and efficiently controlling the input of water, nutrients, and other materials at the right growth stages.[19]

Greengrass shows up as a potent instrument for quickly and effectively updating cities to become smart cities. Greengrass gathers and processes data at the edge before storing it in the cloud thanks to its flexible and configurable sensor nodes that are wisely placed throughout a city. The enhanced capabilities of Greengrass are beneficial in certain fields, including video surveillance, smart transportation, online healthcare, bee colony surveillance, ecological surveillance, and city environment management. The examined data are frequently made available to the public as free of charge resources and are important assets for cutting-edge analytical study.[7][13]

With the number of vehicles on the road increasing daily, ensuring smooth traffic flow in the future demands a stable and scalable system. Present IoT-based vehicle tracking and traffic control systems heavily rely on cloud-based solutions, making their efficiency dependent on bandwidth availability and low latency networks. Currently, 5G networks are limited in coverage, while 2G networks are more widespread. Vehicular IoTs often use 2G for communication between IoT devices and the cloud, but the 64Kbps bandwidth limitation poses challenges. Implementing AWS Greengrass can significantly impact bandwidth usage. The edge device efficiently preprocesses collected data locally by compressing it into a certain format before sending it to the cloud. Autonomous vehicles, requiring real-time data processing to prevent life-threatening accidents, necessitate a shift from cloud to edge computing for computing and processing needs. By embracing AWS Greengrass, the transition to edge processing becomes a pivotal step in advancing vehicular IoTs and supporting safer, faster, and more reliable transportation systems.[9][12]

Enhancing the security of IT systems, both at home and in businesses, is of utmost importance. AWS Greengrass service offers a valuable solution to construct a hack-proof and dynamic security system for IT infrastructure. The core device plays a crucial role in granting zone-specific rights to users attempting to log into the IT system. To ensure the identification of genuine client computer systems, PCs are equipped with environmental sensors, incorporating parameters such as location, temperature, pressure, light, images, and sound. These realistic position parameters hold greater efficacy compared to traditional methods like user ID & password, biometrics, tokens, or patterns for safeguarding IT systems against potential hacks, especially in 3D zones and beyond. [8]

## 5. CONCLUSION

This paper covered the definition of edge computing. AWS Greengrass was introduced, its advantages and limitations were reviewed, but more importantly, its impact on numerous disciplines was examined, and inferenced that it has enormous potential in many different industrial sectors.